\documentclass[allclo]{FBSart}
\usepackage{amsfonts}
\usepackage{amssymb}
\usepackage{epsfig}

\title{The neutron transversity from semi-inclusive DIS off
$^3$He}
\author{S.\,Scopetta}
\institute{Dipartimento di Fisica, Universit\`a degli Studi
di Perugia, \\
and INFN, sezione di Perugia, via A. Pascoli,
06100 Perugia, Italy}

\runningauthor{S.\,Scopetta}
\runningtitle{The neutron transversity from semi-inclusive DIS off
$^3$He}
\begin{document}

\maketitle
\begin{abstract}
A calculation of nuclear effects in the extraction of neutron
single spin asymmetries in semi-inclusive deep inelastic scattering
off $^3$He is described. In the kinematics of forth-coming
experiments at JLab, it is found that the nuclear effects arising within an 
Impulse Approximation approach are under control.
\end{abstract}

A puzzling
experimental scenario has arisen recently,
after measurements of semi-inclusive deep inelastic scattering
(SIDIS) off transversely polarized
proton and deuteron targets \cite{data}. The data show
a strong, unexpected flavour dependence in the azimuthal
distribution of the produced pions. 
With the aim at extracting the neutron information
to shed some light on the problem,
a measurement of SIDIS
off transversely polarized $^3$He has been addressed \cite{bro},
and two experiments, planned to measure azimuthal asymmetries
in the production of leading $\pi^\pm$  from transversely
polarized $^3$He, are forth-coming at JLab \cite{ceb}.
Here, a recent, realistic analysis of SIDIS
off transversely polarized $^3$He \cite{mio} is described.
The formal expressions
of the
Collins and Sivers contributions to the azimuthal
Single Spin Asymmetry (SSA) for the production
of leading pions have been derived, in impulse approximation (IA),
including also the initial transverse momentum of the struck quark.
The final equations are rather involved and they are not
reported here. They can be found in \cite{mio}.
The same quantities have been then evaluated
in the kinematics of the planned 
JLab experiments.
Wave functions \cite{pisa} obtained within
the AV18 interaction \cite{av18} have been used for a realistic
description of the nuclear dynamics,
using overlap integrals evaluated in Ref. \cite{over},
and the nucleon structure has been described
by proper parameterizations of data or suitable model calculations
\cite{model}.
\begin{figure}[thb]
  \hfill
  \begin{minipage}[t]{.45\textwidth}
    \begin{center}  
      \epsfig{file=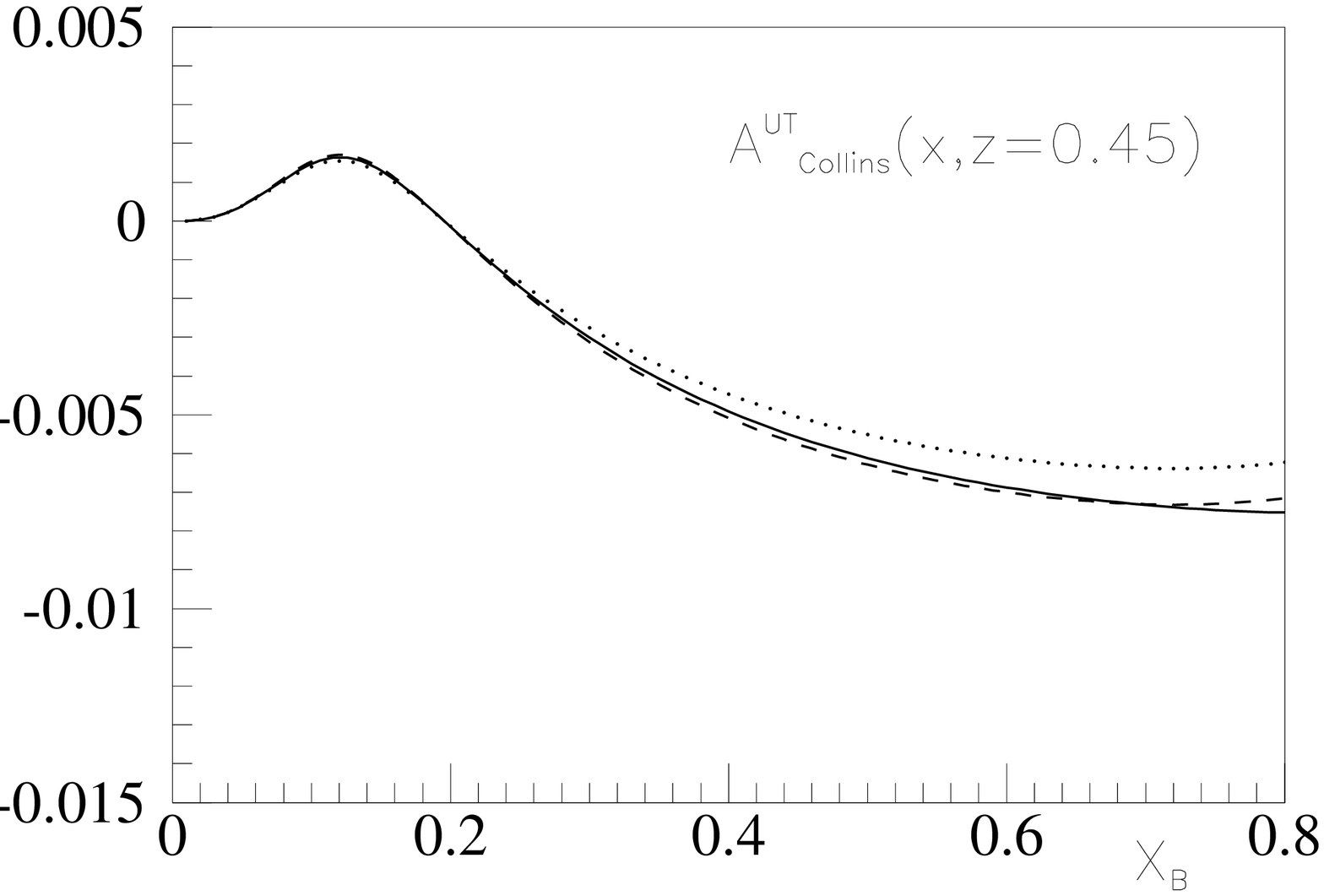, width=6.5cm}
      \vspace{-5cm}
      \caption{The model neutron Collins asymmetry for
$\pi^-$ production
(full) in JLab kinematics, and the one extracted
from the full calculation taking into account
the proton effective polarization
(dashed), or neglecting it (dotted).} 
      \label{fig1a-tc}
    \end{center}
  \end{minipage}
  \hfill
  \begin{minipage}[t]{.45\textwidth}
    \begin{center}  
      \epsfig{file=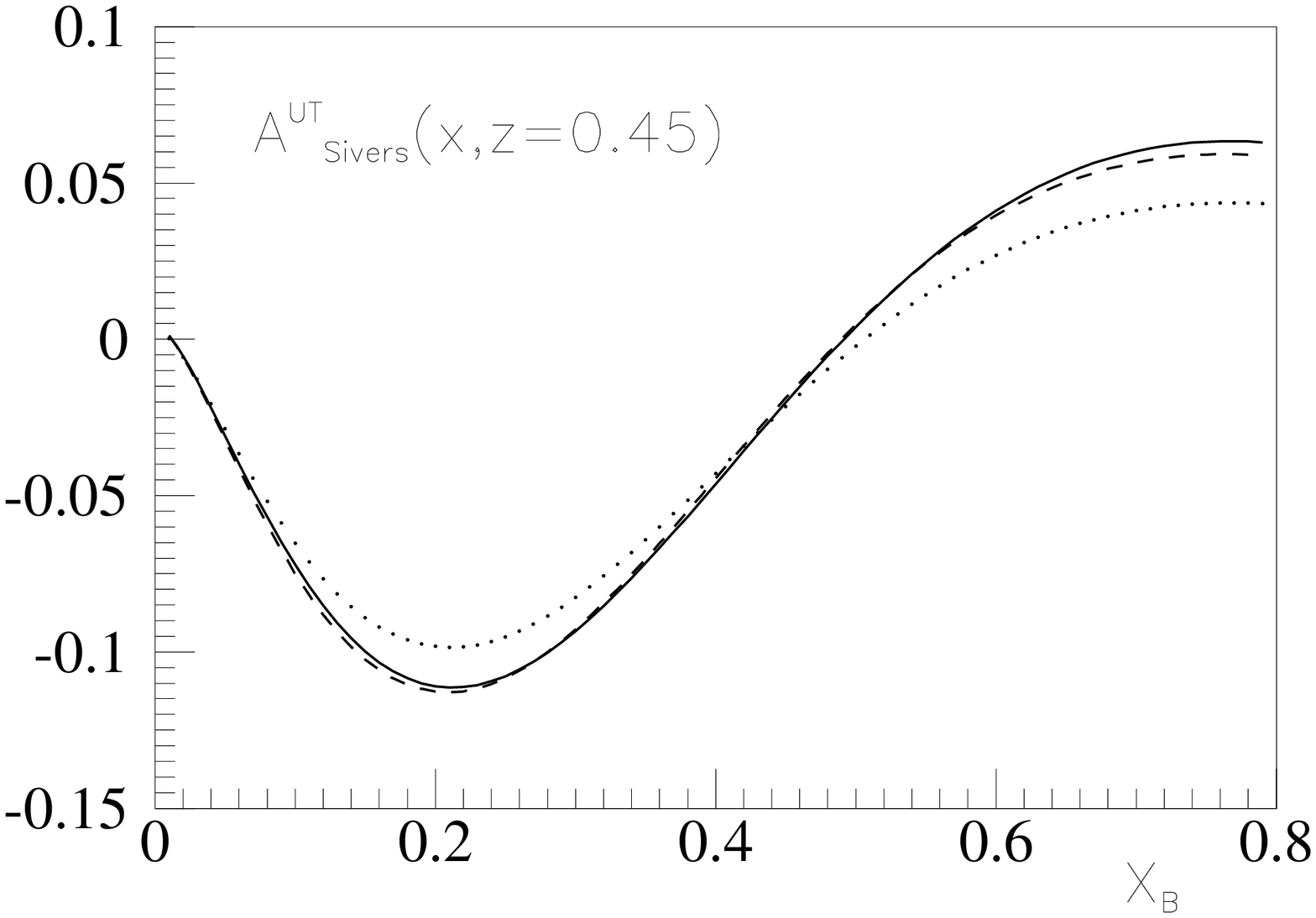, width=6.5cm}
      \vspace{-5cm}
      \caption{The same as in Fig. 1, but for the Sivers asymmetry.
               The results are shown for {$z$}=0.45 and
$Q^2= 2.2$ GeV$^2$, typical values
               in the kinematics of the JLab experiments.}
      \label{fig1b-tc}
    \end{center}
  \end{minipage}
  \hfill
\end{figure}
The crucial issue of extracting
the neutron information from $^3$He data
will be now thoroughly discussed. 
As a matter of facts,
a model independent procedure, based
on the realistic evaluation
of the proton and neutron polarizations in $^3$He
\cite{old}, called respectively $p_p$ and $p_n$ in the following,
is widely used in inclusive 
DIS to take into account effectively  
the momentum and energy distributions
of the bound nucleons in $^3$He.
It is found that the same extraction technique
can be applied also in the 
kinematics of the proposed experiments, although
fragmentation functions, not only parton
distributions, are involved, as it can be seen
in Figs. 1 and 2. In these figures,
the free neutron asymmetry used as a model in the
calculation, given by a full line, is compared with two other quantities.
One is:
\begin{equation}
\bar A^i_n \simeq {1 \over d_n} A^{exp,i}_3~,
\label{extr-1}
\end{equation}
where $i$ stands for ``Collins'' or ``Sivers'',
$A^{exp,i}_3$ is the result of the full calculation, 
simulating data, and $d_n$ is the neutron dilution
factor. The latter quantity is defined as follows, for a neutron $n$
(proton $p$) in $^3$He:
\begin{eqnarray}
d_{n(p)}(x,z)=
{\sum_q e_q^2
f^{q,{n(p)}} 
\left ( x \right )
D^{q,h} \left ( z  \right )
\over
\sum_{N=p,n}
\sum_q e_q^2
f^{q,N} 
( x )
D^{q,h} 
\left ( z \right )
}
\label{dilut}
\end{eqnarray}
and, depending on the standard parton
distributions, $ f^{q,N} ( x )$,
and fragmentation functions, $D^{q,h} 
\left ( z \right )$,
is experimentally known (see \cite{mio} for details). 
$\bar A^i_n $ is given by the dotted curve in the figures.
The third curve, the dashed one, is given by 
\begin{equation}
A^i_n \simeq {1 \over p_n d_n} \left ( A^{exp,i}_3 - 2 p_p d_p
A^{exp,i}_p \right )~,
\label{extr}
\end{equation}
i.e. $^3$He is treated as a nucleus
where the effects of its complicated
spin structure, leading to a depolarization
of the bound neutron, together with the ones of
Fermi motion and binding, can be taken care
of by parameterizing the nucleon effective polarizations,
$p_p$ and $p_n$.
One should realize that
Eq. (\ref{extr-1}) is the relation which should hold
between the $^3$He and the neutron SSAs if there were no nuclear effects,
i.e. the $^3$He nucleus were a system of free nucleons in a pure $S$ wave.
In fact, Eq. (\ref{extr-1}) can  be obtained from Eq. (\ref{extr}) by 
imposing $p_n=1$ and $p_p=0$.
It is clear from the figures that the difference 
between the full and dotted curves,
showing the amount of nuclear effects, is sizable,
being around 10 - 15 \% for any experimentally relevant $x$ and $z$,
while the difference between the dashed
and full curves reduces drastically
to a few percent, showing that the extraction
scheme Eq. (\ref{extr}) takes safely into account
the spin structure of $^3$He, together with Fermi
motion and binding effects. 
This important result is due to the peculiar kinematics
of the JLab experiments, which helps in two ways.
First of all, to favor pions from current fragmentation, 
$z$ has been chosen in the range $0.45\lesssim z \lesssim 0.6$,
which means that only high-energy pions are observed.
Secondly, the pions are detected in a narrow cone around the direction
of the momentum transfer. As it is explained in \cite{mio},
this makes nuclear effects in the fragmentation 
functions rather small. The leading nuclear effects are then 
the ones affecting the parton distributions, already found
in inclusive DIS, and can be taken into account
in the usual way, i.e., using Eq. (\ref{extr}) for the extraction of the
neutron information. In the figures,
one should not take the 
shape and size of the asymmetries too seriously,
being the obtained quantities 
strongly dependent on the models chosen for the unknown distributions
\cite{model}.
One should instead consider the difference between
the curves, a model independent
feature which is the most relevant outcome of the present
investigation. 
The main conclusion is that Eq. (\ref{extr}) will be a valuable tool
for the data analysis of
the experiments \cite{ceb}.

The evaluation of possible effects beyond IA, such as 
final state interactions, and the inclusion in the scheme
of more realistic models of the nucleon structure, able to
predict reasonable figures for the experiments, are in progress.

\end{document}